\documentclass[journal,10pt]{IEEEtran}

\usepackage{cite}
\usepackage{graphicx}
\usepackage{amssymb}
\usepackage{amsmath}
\usepackage{amsthm}
\usepackage{booktabs} %\toprule \midrule \bottomrule
\usepackage{algorithm}
\usepackage{algorithmic}
\usepackage{microtype} 
\usepackage{balance}
\usepackage{grffile} %%%% LEAVE THAT COMMAND BC WE HAVE DOUBLE SUFFIX .fig.pdf
\usepackage[dvipsnames]{xcolor}
\usepackage{mathabx}
\usepackage{scalerel}

\usepackage[caption=false,font=footnotesize]{subfig}

\usepackage{amssymb}
\usepackage{amsfonts}
\usepackage{mathrsfs}
\usepackage{xspace}
\usepackage{bm}
\usepackage{upgreek}

\newcommand{\safemath}[2]{\newcommand{#1}{\ensuremath{#2}\xspace}}

\safemath{\bma}{\mathbf{a}}
\safemath{\bmb}{\mathbf{b}}
\safemath{\bmc}{\mathbf{c}}
\safemath{\bmd}{\mathbf{d}}
\safemath{\bme}{\mathbf{e}}
\safemath{\bmf}{\mathbf{f}}
\safemath{\bmg}{\mathbf{g}}
\safemath{\bmh}{\mathbf{h}}
\safemath{\bmi}{\mathbf{i}}
\safemath{\bmj}{\mathbf{j}}
\safemath{\bmk}{\mathbf{k}}
\safemath{\bml}{\mathbf{l}}
\safemath{\bmm}{\mathbf{m}}
\safemath{\bmn}{\mathbf{n}}
\safemath{\bmo}{\mathbf{o}}
\safemath{\bmp}{\mathbf{p}}
\safemath{\bmq}{\mathbf{q}}
\safemath{\bmr}{\mathbf{r}}
\safemath{\bms}{\mathbf{s}}
\safemath{\bmt}{\mathbf{t}}
\safemath{\bmu}{\mathbf{u}}
\safemath{\bmv}{\mathbf{v}}
\safemath{\bmw}{\mathbf{w}}
\safemath{\bmx}{\mathbf{x}}
\safemath{\bmy}{\mathbf{y}}
\safemath{\bmz}{\mathbf{z}}
\safemath{\bmzero}{\mathbf{0}}
\safemath{\bmone}{\mathbf{1}}

\bmdefine{\biad}{a}
\bmdefine{\bibd}{b}
\bmdefine{\bicd}{c}
\bmdefine{\bidd}{d}
\bmdefine{\bied}{e}
\bmdefine{\bifd}{f}
\bmdefine{\bigd}{g}
\bmdefine{\bihd}{h}
\bmdefine{\biid}{i}
\bmdefine{\bijd}{j}
\bmdefine{\bikd}{k}
\bmdefine{\bild}{l}
\bmdefine{\bimd}{m}
\bmdefine{\bind}{n}
\bmdefine{\biod}{o}
\bmdefine{\bipd}{p}
\bmdefine{\biqd}{q}
\bmdefine{\bird}{r}
\bmdefine{\bisd}{s}
\bmdefine{\bitd}{t}
\bmdefine{\biud}{u}
\bmdefine{\bivd}{v}
\bmdefine{\biwd}{w}
\bmdefine{\bixd}{x}
\bmdefine{\biyd}{y}
\bmdefine{\bizd}{z}

\bmdefine{\bixid}{\xi}
\bmdefine{\bilambdad}{\lambda}
\bmdefine{\bimud}{\mu}
\bmdefine{\bithetad}{\theta}
\bmdefine{\biphid}{\phi}
\bmdefine{\bideltad}{\delta}

\safemath{\bmia}{\biad}
\safemath{\bmib}{\bibd}
\safemath{\bmic}{\bicd}
\safemath{\bmid}{\bidd}
\safemath{\bmie}{\bied}
\safemath{\bmif}{\bifd}
\safemath{\bmig}{\bigd}
\safemath{\bmih}{\bihd}
\safemath{\bmii}{\biid}
\safemath{\bmij}{\bijd}
\safemath{\bmik}{\bikd}
\safemath{\bmil}{\bild}
\safemath{\bmim}{\bimd}
\safemath{\bmin}{\bind}
\safemath{\bmio}{\biod}
\safemath{\bmip}{\bipd}
\safemath{\bmiq}{\biqd}
\safemath{\bmir}{\bird}
\safemath{\bmis}{\bisd}
\safemath{\bmit}{\bitd}
\safemath{\bmiu}{\biud}
\safemath{\bmiv}{\bivd}
\safemath{\bmiw}{\biwd}
\safemath{\bmix}{\bixd}
\safemath{\bmiy}{\biyd}
\safemath{\bmiz}{\bizd}

\safemath{\bmxi}{\bixid}
\safemath{\bmlambda}{\bilambdad}
\safemath{\bmmu}{\bimud}
\safemath{\bmtheta}{\bithetad}
\safemath{\bmphi}{\biphid}
\safemath{\bmdelta}{\bideltad}

\safemath{\bA}{\mathbf{A}}
\safemath{\bB}{\mathbf{B}}
\safemath{\bC}{\mathbf{C}}
\safemath{\bD}{\mathbf{D}}
\safemath{\bE}{\mathbf{E}}
\safemath{\bF}{\mathbf{F}}
\safemath{\bG}{\mathbf{G}}
\safemath{\bH}{\mathbf{H}}
\safemath{\bI}{\mathbf{I}}
\safemath{\bJ}{\mathbf{J}}
\safemath{\bK}{\mathbf{K}}
\safemath{\bL}{\mathbf{L}}
\safemath{\bM}{\mathbf{M}}
\safemath{\bN}{\mathbf{N}}
\safemath{\bO}{\mathbf{O}}
\safemath{\bP}{\mathbf{P}}
\safemath{\bQ}{\mathbf{Q}}
\safemath{\bR}{\mathbf{R}}
\safemath{\bS}{\mathbf{S}}
\safemath{\bT}{\mathbf{T}}
\safemath{\bU}{\mathbf{U}}
\safemath{\bV}{\mathbf{V}}
\safemath{\bW}{\mathbf{W}}
\safemath{\bX}{\mathbf{X}}
\safemath{\bY}{\mathbf{Y}}
\safemath{\bZ}{\mathbf{Z}}

\safemath{\bZero}{\mathbf{0}}
\safemath{\bOne}{\mathbf{1}}
\safemath{\bDelta}{\mathbf{\Delta}}
\safemath{\bLambda}{\mathbf{\UpLambda}}
\safemath{\bPhi}{\mathbf{\Phi}}
\safemath{\bPsi}{\mathbf{\Psi}}
\safemath{\bSigma}{\mathbf{\Upsigma}}
\safemath{\bOmega}{\mathbf{\Upomega}}
\safemath{\bTheta}{\mathbf{\Uptheta}}

\bmdefine{\biAd}{A}
\bmdefine{\biBd}{B}
\bmdefine{\biCd}{C}
\bmdefine{\biDd}{D}
\bmdefine{\biEd}{E}
\bmdefine{\biFd}{F}
\bmdefine{\biGd}{G}
\bmdefine{\biHd}{H}
\bmdefine{\biId}{I}
\bmdefine{\biJd}{J}
\bmdefine{\biKd}{K}
\bmdefine{\biLd}{L}
\bmdefine{\biMd}{M}
\bmdefine{\biOd}{N}
\bmdefine{\biPd}{O}
\bmdefine{\biQd}{P}
\bmdefine{\biRd}{R}
\bmdefine{\biSd}{S}
\bmdefine{\biTd}{T}
\bmdefine{\biUd}{U}
\bmdefine{\biVd}{V}
\bmdefine{\biWd}{W}
\bmdefine{\biXd}{X}
\bmdefine{\biYd}{Y}
\bmdefine{\biZd}{Z}

\bmdefine{\biDelta}{\Delta}
\bmdefine{\biLambda}{\Lambda}
\bmdefine{\biPhi}{\Phi}
\bmdefine{\biSigma}{\Sigma}
\bmdefine{\biOmega}{\Omega}
\bmdefine{\biTheta}{\Theta}

\safemath{\bimA}{\biAd}
\safemath{\bimB}{\biBd}
\safemath{\bimC}{\biCd}
\safemath{\bimD}{\biDd}
\safemath{\bimE}{\biEd}
\safemath{\bimF}{\biFd}
\safemath{\bimG}{\biGd}
\safemath{\bimH}{\biHd}
\safemath{\bimI}{\biId}
\safemath{\bimJ}{\biJd}
\safemath{\bimK}{\biKd}
\safemath{\bimL}{\biLd}
\safemath{\bimM}{\biMd}
\safemath{\bimN}{\biNd}
\safemath{\bimO}{\biOd}
\safemath{\bimP}{\biPd}
\safemath{\bimQ}{\biQd}
\safemath{\bimR}{\biRd}
\safemath{\bimS}{\biSd}
\safemath{\bimT}{\biTd}
\safemath{\bimU}{\biUd}
\safemath{\bimV}{\biVd}
\safemath{\bimW}{\biWd}
\safemath{\bimX}{\biXd}
\safemath{\bimY}{\biYd}
\safemath{\bimZ}{\biZd}

\safemath{\bimDelta}{\biDelta}
\safemath{\bimLambda}{\biLambda}
\safemath{\bimPhi}{\biPhi}
\safemath{\bimSigma}{\biSigma}
\safemath{\bimOmega}{\biOmega}
\safemath{\bimTheta}{\biTheta}

\safemath{\setA}{\mathcal{A}}
\safemath{\setB}{\mathcal{B}}
\safemath{\setC}{\mathcal{C}}
\safemath{\setD}{\mathcal{D}}
\safemath{\setE}{\mathcal{E}}
\safemath{\setF}{\mathcal{F}}
\safemath{\setG}{\mathcal{G}}
\safemath{\setH}{\mathcal{H}}
\safemath{\setI}{\mathcal{I}}
\safemath{\setJ}{\mathcal{J}}
\safemath{\setK}{\mathcal{K}}
\safemath{\setL}{\mathcal{L}}
\safemath{\setM}{\mathcal{M}}
\safemath{\setN}{\mathcal{N}}
\safemath{\setO}{\mathcal{O}}
\safemath{\setP}{\mathcal{P}}
\safemath{\setQ}{\mathcal{Q}}
\safemath{\setR}{\mathcal{R}}
\safemath{\setS}{\mathcal{S}}
\safemath{\setT}{\mathcal{T}}
\safemath{\setU}{\mathcal{U}}
\safemath{\setV}{\mathcal{V}}
\safemath{\setW}{\mathcal{W}}
\safemath{\setX}{\mathcal{X}}
\safemath{\setY}{\mathcal{Y}}
\safemath{\setZ}{\mathcal{Z}}
\safemath{\emptySet}{\varnothing}

\safemath{\colA}{\mathscr{A}}
\safemath{\colB}{\mathscr{B}}
\safemath{\colC}{\mathscr{C}}
\safemath{\colD}{\mathscr{D}}
\safemath{\colE}{\mathscr{E}}
\safemath{\colF}{\mathscr{F}}
\safemath{\colG}{\mathscr{G}}
\safemath{\colH}{\mathscr{H}}
\safemath{\colI}{\mathscr{I}}
\safemath{\colJ}{\mathscr{J}}
\safemath{\colK}{\mathscr{K}}
\safemath{\colL}{\mathscr{L}}
\safemath{\colM}{\mathscr{M}}
\safemath{\colN}{\mathscr{N}}
\safemath{\colO}{\mathscr{O}}
\safemath{\colP}{\mathscr{P}}
\safemath{\colQ}{\mathscr{Q}}
\safemath{\colR}{\mathscr{R}}
\safemath{\colS}{\mathscr{S}}
\safemath{\colT}{\mathscr{T}}
\safemath{\colU}{\mathscr{U}}
\safemath{\colV}{\mathscr{V}}
\safemath{\colW}{\mathscr{W}}
\safemath{\colX}{\mathscr{X}}
\safemath{\colY}{\mathscr{Y}}
\safemath{\colZ}{\mathscr{Z}}

\safemath{\opA}{\mathbb{A}}
\safemath{\opB}{\mathbb{B}}
\safemath{\opC}{\mathbb{C}}
\safemath{\opD}{\mathbb{D}}
\safemath{\opE}{\mathbb{E}}
\safemath{\opF}{\mathbb{F}}
\safemath{\opG}{\mathbb{G}}
\safemath{\opH}{\mathbb{H}}
\safemath{\opI}{\mathbb{I}}
\safemath{\opJ}{\mathbb{J}}
\safemath{\opK}{\mathbb{K}}
\safemath{\opL}{\mathbb{L}}
\safemath{\opM}{\mathbb{M}}
\safemath{\opN}{\mathbb{N}}
\safemath{\opO}{\mathbb{O}}
\safemath{\opP}{\mathbb{P}}
\safemath{\opQ}{\mathbb{Q}}
\safemath{\opR}{\mathbb{R}}
\safemath{\opS}{\mathbb{S}}
\safemath{\opT}{\mathbb{T}}
\safemath{\opU}{\mathbb{U}}
\safemath{\opV}{\mathbb{V}}
\safemath{\opW}{\mathbb{W}}
\safemath{\opX}{\mathbb{X}}
\safemath{\opY}{\mathbb{Y}}
\safemath{\opZ}{\mathbb{Z}}
\safemath{\opZero}{\mathbb{O}}
\safemath{\identityop}{\opI}

\safemath{\veca}{\bma}
\safemath{\vecb}{\bmb}
\safemath{\vecc}{\bmc}
\safemath{\vecd}{\bmd}
\safemath{\vece}{\bme}
\safemath{\vecf}{\bmf}
\safemath{\vecg}{\bmg}
\safemath{\vech}{\bmh}
\safemath{\veci}{\bmi}
\safemath{\vecj}{\bmj}
\safemath{\veck}{\bmk}
\safemath{\vecl}{\bml}
\safemath{\vecm}{\bmm}
\safemath{\vecn}{\bmn}
\safemath{\veco}{\bmo}
\safemath{\vecp}{\bmp}
\safemath{\vecq}{\bmq}
\safemath{\vecr}{\bmr}
\safemath{\vecs}{\bms}
\safemath{\vect}{\bmt}
\safemath{\vecu}{\bmu}
\safemath{\vecv}{\bmv}
\safemath{\vecw}{\bmw}
\safemath{\vecx}{\bmx}
\safemath{\vecy}{\bmy}
\safemath{\vecz}{\bmz}

\safemath{\veczero}{\bmzero}
\safemath{\vecone}{\bmone}
\safemath{\vecxi}{\bmxi}
\safemath{\veclambda}{\bmlambda}
\safemath{\vecmu}{\bmmu}
\safemath{\vectheta}{\bmtheta}
\safemath{\vecphi}{\bmphi}
\safemath{\vecdelta}{\bmdelta}

\safemath{\matA}{\bA}
\safemath{\matB}{\bB}
\safemath{\matC}{\bC}
\safemath{\matD}{\bD}
\safemath{\matE}{\bE}
\safemath{\matF}{\bF}
\safemath{\matG}{\bG}
\safemath{\matH}{\bH}
\safemath{\matI}{\bI}
\safemath{\matJ}{\bJ}
\safemath{\matK}{\bK}
\safemath{\matL}{\bL}
\safemath{\matM}{\bM}
\safemath{\matN}{\bN}
\safemath{\matO}{\bO}
\safemath{\matP}{\bP}
\safemath{\matQ}{\bQ}
\safemath{\matR}{\bR}
\safemath{\matS}{\bS}
\safemath{\matT}{\bT}
\safemath{\matU}{\bU}
\safemath{\matV}{\bV}
\safemath{\matW}{\bW}
\safemath{\matX}{\bX}
\safemath{\matY}{\bY}
\safemath{\matZ}{\bZ}
\safemath{\matzero}{\bmzero}

\safemath{\matDelta}{\bDelta}
\safemath{\matLambda}{\bLambda}
\safemath{\matPhi}{\bPhi}
\safemath{\matSigma}{\bSigma}
\safemath{\matOmega}{\bOmega}
\safemath{\matTheta}{\bTheta}

\safemath{\matidentity}{\matI}
\safemath{\matone}{\matO}

\safemath{\rnda}{A}
\safemath{\rndb}{B}
\safemath{\rndc}{C}
\safemath{\rndd}{D}
\safemath{\rnde}{E}
\safemath{\rndf}{F}
\safemath{\rndg}{G}
\safemath{\rndh}{H}
\safemath{\rndi}{I}
\safemath{\rndj}{J}
\safemath{\rndk}{K}
\safemath{\rndl}{L}
\safemath{\rndm}{M}
\safemath{\rndn}{N}
\safemath{\rndo}{O}
\safemath{\rndp}{P}
\safemath{\rndq}{Q}
\safemath{\rndr}{R}
\safemath{\rnds}{S}
\safemath{\rndt}{T}
\safemath{\rndu}{U}
\safemath{\rndv}{V}
\safemath{\rndw}{W}
\safemath{\rndx}{X}
\safemath{\rndy}{Y}
\safemath{\rndz}{Z}

\safemath{\rveca}{\bimA}
\safemath{\rvecb}{\bimB}
\safemath{\rvecc}{\bimC}
\safemath{\rvecd}{\bimD}
\safemath{\rvece}{\bimE}
\safemath{\rvecf}{\bimF}
\safemath{\rvecg}{\bimG}
\safemath{\rvech}{\bimH}
\safemath{\rveci}{\bimI}
\safemath{\rvecj}{\bimJ}
\safemath{\rveck}{\bimK}
\safemath{\rvecl}{\bimL}
\safemath{\rvecm}{\bimM}
\safemath{\rvecn}{\bimN}
\safemath{\rveco}{\bomO}
\safemath{\rvecp}{\bimP}
\safemath{\rvecq}{\bimQ}
\safemath{\rvecr}{\bimR}
\safemath{\rvecs}{\bimS}
\safemath{\rvect}{\bimT}
\safemath{\rvecu}{\bimU}
\safemath{\rvecv}{\bimV}
\safemath{\rvecw}{\bimW}
\safemath{\rvecx}{\bimX}
\safemath{\rvecy}{\bimY}
\safemath{\rvecz}{\bimZ}

\safemath{\rvecxi}{\bmxi}
\safemath{\rveclambda}{\bmlambda}
\safemath{\rvecmu}{\bmmu}
\safemath{\rvectheta}{\bmtheta}
\safemath{\rvecphi}{\bmphi}

\safemath{\rmatA}{\bimA}
\safemath{\rmatB}{\bimB}
\safemath{\rmatC}{\bimC}
\safemath{\rmatD}{\bimD}
\safemath{\rmatE}{\bimE}
\safemath{\rmatF}{\bimF}
\safemath{\rmatG}{\bimG}
\safemath{\rmatH}{\bimH}
\safemath{\rmatI}{\bimI}
\safemath{\rmatJ}{\bimJ}
\safemath{\rmatK}{\bimK}
\safemath{\rmatL}{\bimL}
\safemath{\rmatM}{\bimM}
\safemath{\rmatN}{\bimN}
\safemath{\rmatO}{\bimO}
\safemath{\rmatP}{\bimP}
\safemath{\rmatQ}{\bimQ}
\safemath{\rmatR}{\bimR}
\safemath{\rmatS}{\bimS}
\safemath{\rmatT}{\bimT}
\safemath{\rmatU}{\bimU}
\safemath{\rmatV}{\bimV}
\safemath{\rmatW}{\bimW}
\safemath{\rmatX}{\bimX}
\safemath{\rmatY}{\bimY}
\safemath{\rmatZ}{\bimZ}

\safemath{\rmatDelta}{\bimDelta}
\safemath{\rmatLambda}{\bimLambda}
\safemath{\rmatPhi}{\bimPhi}
\safemath{\rmatSigma}{\bimSigma}
\safemath{\rmatOmega}{\bimOmega}
\safemath{\rmatTheta}{\bimTheta}

\usepackage{amssymb}
\usepackage{amsfonts}
\usepackage{mathrsfs}
\usepackage{xspace}
\usepackage{bm}
\usepackage{fancyref}
\usepackage{textcomp}

\usepackage{multirow}
\usepackage{stmaryrd}

\newenvironment{textbmatrix}{	\setlength{\arraycolsep}{2.5pt}%
								\big[\begin{matrix}}{\end{matrix}\big]%
								\raisebox{0.08ex}{\vphantom{M}}}

\def\be{\begin{equation}}
\def\ee{\end{equation}}
\def\een{\nonumber \end{equation}}
\def\mat{\begin{bmatrix}}
\def\emat{\end{bmatrix}}
\def\btm{\begin{textbmatrix}}
\def\etm{\end{textbmatrix}}

\def\ba#1\ea{\begin{align}#1\end{align}}
\def\bas#1\eas{\begin{align*}#1\end{align*}}
\def\bs#1\es{\begin{split}#1\end{split}} 
\def\bg#1\eg{\begin{gather}#1\end{gather}}
\def\bml#1\eml{\begin{multline}#1\end{multline}}
\def\bi#1\ei{\begin{itemize}#1\end{itemize}}

\DeclareMathOperator{\tr}{tr}				% trace
\DeclareMathOperator*{\argmin}{arg\;min}		% arg min
\DeclareMathOperator*{\argmax}{arg\;max}		% arg max
\newcommand{\Ex}[2]{\ensuremath{\Exop_{#1}\lefto[#2\right]}} 	% expectation
\safemath{\dirac}{\delta}					% Dirac delta
\safemath{\krond}{\dirac}					% Kronecker delta
\safemath{\upto}{\uparrow}
\safemath{\downto}{\downarrow}
\safemath{\iu}{j}							% imaginary unit
\safemath{\ev}{\lambda}						% eigenvalue
\safemath{\hilseqspace}{l^{2}}				% Hilbert sequence space
\newcommand{\banachfunspace}[1]{\setL^{#1}}	% Banach function space
\safemath{\hilfunspace}{\banachfunspace{2}}	% Hilbert function space
\safemath{\SNR}{\text{\sc snr}} 				% signal to noise ratio
\safemath{\No}{N_0}							% noise spectral density
\safemath{\Es}{E_s}							% energy per symbol
\safemath{\Eb}{E_b}							% energy per bit
\safemath{\EbNo}{\frac{\Eb}{\No}}
\safemath{\EsNo}{\frac{\Es}{\No}}

\DeclareMathOperator{\CHop}{\ensuremath{\opH}} % channel operator
\safemath{\tvir}{\rndh_{\CHop}}				% time-varying impulse response
\safemath{\tvtf}{\rndl_{\CHop}}				% 	-''- transfer function
\safemath{\spf}{\rnds_{\CHop}}				% spreading function
\safemath{\bff}{H_{\CHop}}					% bi-freuqency function

\safemath{\ircf}{r_{h}}						% impulse response correlation fn.
\safemath{\tftvcf}{r_{s}}					% scattering function
\safemath{\tfcf}{r_{l}}						% time-frequency correlation fn.
\safemath{\bfcf}{r_{H}}						% bi-frequency correlation fn.

\safemath{\tcorr}{c_h}						% time-correlation function
\safemath{\scf}{c_{s}}						% spreading function
\safemath{\tfcorr}{c_{l}}					% transfer-function correlation
\safemath{\fcorr}{c_{H}}						% frequency-correlation function

\safemath{\mi}{I}							% mutual information
\safemath{\capacity}{C}						% capacity

\safemath{\normal}{\mathcal{N}}			% normal distribution
\safemath{\jpg}{\mathcal{CN}}			% jointly proper Gaussian
\safemath{\mchain}{\leftrightarrow}		% Markov chain
\safemath{\dB}{\,\mathrm{dB}}
\safemath{\dBm}{\,\mathrm{dBm}}
\safemath{\Hz}{\,\mathrm{Hz}}
\safemath{\kHz}{\,\mathrm{kHz}}
\safemath{\MHz}{\,\mathrm{MHz}}
\safemath{\GHz}{\,\mathrm{GHz}}
\safemath{\s}{\,\mathrm{s}}
\safemath{\ms}{\,\mathrm{ms}}
\safemath{\mus}{\,\mathrm{\text{\textmu}s}}
\safemath{\ns}{\,\mathrm{ns}}
\safemath{\ps}{\,\mathrm{ps}}
\safemath{\meter}{\,\mathrm{m}}
\safemath{\mm}{\,\mathrm{mm}}
\safemath{\cm}{\,\mathrm{cm}}
\safemath{\W}{\,\mathrm{W}}
\safemath{\mW}{\, \mathrm{mW}}
\safemath{\J}{\,\mathrm{J}}
\safemath{\K}{\,\mathrm{K}}
\safemath{\bit}{\,\mathrm{bit}}
\safemath{\nat}{\,\mathrm{nat}}

\safemath{\define}{\triangleq}			% definition

\safemath{\equivalent}{\sim}
\safemath{\distas}{\sim}					% distributed according to
\safemath{\sdiff}{\Delta}				% symmetric set difference

\safemath{\reals}{\mathbb{R}}
\safemath{\positivereals}{\reals_{+}}
\safemath{\integers}{\mathbb{Z}}
\safemath{\posint}{\integers_{+}}
\safemath{\naturals}{\mathbb{N}}
\safemath{\posnaturals}{\naturals_{+}}
\safemath{\complexset}{\mathbb{C}}
\safemath{\rationals}{\mathbb{Q}}

\newcommand*{\fancyrefapplabelprefix}{app}		% Appendix
\newcommand*{\fancyrefthmlabelprefix}{thm}		% Theorem
\newcommand*{\fancyreflemlabelprefix}{lem}		% Lemma
\newcommand*{\fancyrefcorlabelprefix}{cor}		% Corollary
\newcommand*{\fancyrefdeflabelprefix}{def}		% Definition
\newcommand*{\fancyrefproplabelprefix}{prop}		% Property
\newcommand*{\fancyrefexmpllabelprefix}{exmpl}
\newcommand*{\fancyreftbllabelprefix}{tbl}		% Lemma
\frefformat{vario}{\fancyrefseclabelprefix}{Section~#1}
\frefformat{vario}{\fancyrefthmlabelprefix}{Theorem~#1}
\frefformat{vario}{\fancyreflemlabelprefix}{Lemma~#1}
\frefformat{vario}{\fancyrefcorlabelprefix}{Corollary~#1}
\frefformat{vario}{\fancyrefdeflabelprefix}{Definition~#1}
\frefformat{vario}{\fancyreffiglabelprefix}{Fig.~#1}
\frefformat{vario}{\fancyreftbllabelprefix}{Tbl.~#1}
\frefformat{vario}{\fancyrefapplabelprefix}{Appendix~#1}
\frefformat{vario}{\fancyrefeqlabelprefix}{(#1)}
\frefformat{vario}{\fancyrefproplabelprefix}{Property~#1}
\frefformat{vario}{\fancyrefexmpllabelprefix}{Example~#1}

\newcommand{\marginnote}[1]{\marginpar{\framebox{\framebox{#1}}}}
\newcommand{\comment}[2]{[\marginnote{#1}\textit{#1}:\ \textit{#2}]}

\safemath{\ysig}{\bmy}
\safemath{\ysighat}{\hat{\ysig}}
\safemath{\ysigdim}{M}
\safemath{\xsig}{\bmx}
\safemath{\xsigdim}{N}
\safemath{\nx}{n_x}
\safemath{\zsig}{\bmz}
\safemath{\zsigdim}{\ysigdim}
\safemath{\rsig}{\bmr}
\safemath{\Adict}{\bA}
\safemath{\Adicttilde}{\widetilde{\Adict}}
\safemath{\Adictdim}{\outputdim\times\xsigdim}
\safemath{\avec}{\bma}
\safemath{\avectilde}{\tilde{\avec}}
\safemath{\Bdict}{\bB}
\safemath{\Bdicttilde}{\widetilde{\Bdict}}
\safemath{\Cdict}{\bC}
\safemath{\cvec}{\bmc}
\safemath{\Ddict}{\bD}
\safemath{\Ddictdim}{\ysigdim\times\xsigdim}
\safemath{\dvec}{\bmd}
\safemath{\Ddicttilde}{\widetilde{\bD}}
\safemath{\Bonb}{\bB}
\safemath{\bvec}{\bmb}
\safemath{\Bonbdim}{\ysigdim\times\ysigdim}
\safemath{\noise}{\bmn}
\safemath{\noisedim}{\ysigim}
\safemath{\err}{\bme}
\safemath{\errdim}{\ysigdim}
\safemath{\errset}{\setE}
\safemath{\nerr}{n_e}
\safemath{\delop}{\bP_\errset}
\safemath{\delopc}{\bP_{{\errset}^c}}

\safemath{\cplxi}{\imath}
\safemath{\cplxj}{\jmath}
\safemath{\dict}{\matD}
\safemath{\inputdim}{N}		% number of columns of dictionary D
\safemath{\outputdim}{M}		%number of rows of dictionary D
\safemath{\sparsity}{s}	%sparsity
\safemath{\inputdimA}{{N_a}}	%total number of elements in dictionary A
\safemath{\inputdimB}{{N_b}}	%total number of elements in dictionary B
\safemath{\elemA}{{n_a}}	%number of elements chosen from dictionary A
\safemath{\elemB}{{n_b}}	%number of elements chosen from dictionary B
\safemath{\resA}{\matR_a}	%restriction map to elements of dictionary A
\safemath{\resB}{\matR_b}	%restriction map to elements of dictionary B
\safemath{\subD}{\matS} %subdictionary
\safemath{\subA}{\matS_a} %subdictionary part of A
\safemath{\subB}{\matS_b} %subdictionary part of B
\safemath{\dicta}{\matA} 	% first subdictionary
\safemath{\dictb}{\matB} 	% second subdictionary
\safemath{\hollowS}{H}
\safemath{\hollowA}{H_a}
\safemath{\hollowB}{H_b}
\safemath{\cross}{Z}
\safemath{\coh}{\mu}			% coherence dictionary
\safemath{\coha}{\mu_a}			% coherence first subdictionary
\safemath{\cohb}{\mu_b}			% coherence second subdictionary
\safemath{\mubs}{\nu}	%block sub-coherence
\safemath{\cohm}{\mu_m} %mutual coherence
\safemath{\dictset}{\setD}	% set of dictionaries
\safemath{\dictsetp}{\dictset(\coh,\coha,\cohb)}	% set of dictionaries parametrized
\safemath{\dictsetgen}{\dictset_\text{gen}}
\safemath{\dictsetgenp}{\dictsetgen(\coh)}
\safemath{\dictsetonb}{\dictset_\text{onb}}
\safemath{\dictsetonbp}{\dictsetonb(\coh)}

\safemath{\leftside}{U}
\safemath{\rightsideA}{R_a}
\safemath{\rightsideB}{R_b}

\safemath{\indexS}{\setI_S} %set of indices participating in sub-dictionary S

\safemath{\na}{n_a}			% cardinality of set of linearly independent columns of first dictionary
\safemath{\nb}{n_b}			% cardinality of set of linearly independent columns of second dictionary
\safemath{\coeffa}{p_i}	%coefficients for columns of A
\safemath{\coeffb}{q_j}	%coefficients for columns of B
\safemath{\seta}{\setP}		% set of linearly independent columns of A
\safemath{\setb}{\setQ}     % set of linearly independent columns of B
\safemath{\setw}{\setW}	%set of n largest elements of w
\safemath{\setz}{\setZ}	%set of L-n largest elements of z
\safemath{\cola}{\veca}		% generic element of the dictionary A
\safemath{\colb}{\vecb}		% generic element of the dictionary B
\safemath{\cold}{\vecd}		% generic element of the dictionary D
\safemath{\inputvec}{\vecx} 	%coefficient vector (input)
\safemath{\error}{\vece}	%error vector
\safemath{\noiseout}{\vecz} 	%noisy output vector
\safemath{\inputvecel}{x}
\safemath{\inputveca}{\vecx_a}
\safemath{\inputvecb}{\vecx_b}
\safemath{\outputvec}{\vecy}	%output of Dictionary
\safemath{\lambdamin}{\lambda_{\mathrm{min}}}
\safemath{\elltwo}{\ell_2}
\safemath{\ellone}{\ell_1}
\safemath{\ellzero}{\ell_0}
\safemath{\ellinf}{\ell_\infty}
\safemath{\licard}{Z(\coh,\coha,\cohb)}
\safemath{\xsol}{\hat{x}}
\safemath{\xbord}{x_b}		%Solution at the border
\safemath{\xstat}{x_s}		%Solution stationary in l0 prob
\safemath{\xstatLone}{\tilde{x}_s}
\safemath{\order}{\mathcal{O}} %order notation (big O)
\safemath{\scales}{\Theta} %scales as
\safemath{\ones}{\mathbf{1}} %all ones matrix
\safemath{\zeroes}{\mathbf{0}} %all zeroes matrix
\safemath{\thlone}{\kappa(\coh,\cohb)} %treshold l1 problem
\safemath{\constoneA}{\delta} %constant in l1 theorem to save space
\safemath{\constoneB}{\epsilon} %constant in l1 theorem to save space
\safemath{\nlarge}{L}				   %num large elements
\safemath{\sumlarge}{S_\nlarge}
\safemath{\maxlarger}{P_\nlarge}	   % maximum in Gribonval and Nielsen
\safemath{\Pzero}{\textrm{P0}}	
\safemath{\Pone}{\textrm{P1}}
\safemath{\vecfir}{\vecw}			 % \vecv element of the kernel of the dictionary, \vecv=[\vecfir \vecsec]
\safemath{\vecsec}{\vecz}
\safemath{\elvecfir}{w}              % element of vecfir
\safemath{\elvecsec}{z}				 % element of vecsec
\safemath{\nlargefir}{n}
\safemath{\normout}{\gamma}
\safemath{\auxfun}{h}
\safemath{\supp}{\textrm{supp}}%support

\safemath{\indexa}{\ell}
\safemath{\indexb}{r}
\safemath{\indexc}{i}
\safemath{\indexd}{j}

\safemath{\project}{P}%projector

\usepackage{color}

\usepackage{enumitem}
\setlist[itemize]{leftmargin=*, itemsep=0.3em, topsep=0.3em} % makes itemization a bit more compact (no indents)

\allowdisplaybreaks % THIS ALLOWS EQUATIONS TO BREAK THE PAGE

\newcommand{\eg}[1]{\comment{eg}{\bf \textcolor{Fuchsia}{#1}}}

\safemath{\LAMA}{\textrm{LAMA}}
\safemath{\MRT}{\textrm{MRT}}
\safemath{\betamax}{\beta^\text{max}_\setO}
\safemath{\betamaxno}{\beta^\text{max}}
\safemath{\betamin}{\beta^\text{min}_\setO}
\safemath{\betaminno}{\beta^\text{min}}

\safemath{\Nomin}{\No^\textnormal{min}(\beta)}
\safemath{\Nominnobeta}{\No^\text{min}}
\safemath{\Nomax}{\No^\textnormal{max}(\beta)}
\safemath{\Nomaxnobeta}{\No^\textnormal{max}}
\safemath{\EX}{E_\textnormal{x}}
\safemath{\EXP}{\EX^\textnormal{p}}
\safemath{\Eo}{E_0}

\safemath{\tmax}{{t_\textnormal{max}}}
\safemath{\MAP}{\textrm{MAP}}
\safemath{\IO}{\textrm{IO}}
\safemath{\JO}{\textrm{JO}}
\safemath{\Nopost}{N_{0}^\textnormal{post}}
\safemath{\MT}{U}
\safemath{\MR}{B}
\safemath{\Tran}{\textnormal{T}}
\safemath{\Herm}{\textnormal{H}}
\safemath{\row}{\textnormal{r}}
\safemath{\col}{\textnormal{c}}

\safemath{\NT}{N_\textnormal{T}}
\safemath{\DSNR}{\delta \textnormal{SNR}}
\safemath{\betaMOR}{\beta^{\star}}

\usepackage{amssymb}
\newcommand{\SBP}{\text{SBP}}
\def\Omegac{\scalerel*{\rotatebox[origin=c]{180}{$\Omega$}}{\Omega}}
\newcommand{\setint}[1]{\ldbrack #1 \rdbrack}
\newcommand{\udl}[1]{\underline{#1}}
\title{Hardware-Aware Beamspace Precoding for All-Digital mmWave Massive MU-MIMO}
\author{Emre G\"{o}n\"{u}lta\c{s}{$^\star$}, Sueda {Taner}{$^\star$},  Alexandra Gallyas-Sanhueza, Seyed Hadi Mirfarshbafan, and Christoph Studer\thanks{$^\star$EG and ST contributed equally to this work.}\thanks{EG, ST, and AGS are with the School of Electrical and Computer Engineering, Cornell University, Ithaca, NY 14853;  e-mail: eg566@cornell.edu, st939@cornell.edu, ag753@cornell.edu.}
\thanks{SHM and CS are with the Department of Information Technology and Electrical Engineering, ETH Z\"urich, Z\"urich, Switzerland; e-mail: \mbox{mirfarshbafan@iis.ee.ethz.ch}, studer@ethz.ch.}\thanks{The work of ST, AGS, and SHM was supported by  ComSenTer, one of six centers in JUMP, an SRC program sponsored by DARPA. The work of EG and CS was supported by the US NSF grants CNS-1717559 and ECCS-1824379. The work of SHM and CS was also supported by an ETH Research Grant.}\thanks{The authors thank O. Casta\~neda for discussions on computational complexity.}}

\begin{document}

\maketitle

% !TEX root = 21COMLET_beampre.tex
% DO NOT REMOVE THE ABOVE COMMENT!

\begin{abstract}
Massive multi-user multiple-input multiple-output (MU-MIMO) wireless systems operating at millimeter-wave (mmWave) frequencies enable simultaneous wideband data transmission to a large number of users. In order to reduce the complexity of MU precoding in {all-digital basestation architectures}, we propose a two-stage precoding architecture that first performs precoding using a sparse matrix in the beamspace domain, followed by an inverse fast Fourier transform that converts the result to the antenna domain. The sparse precoding matrix requires a small number of multipliers and enables regular hardware architectures, which allows the design of hardware-efficient all-digital precoders. Simulation results demonstrate that our methods approach the error-rate of conventional Wiener filter precoding with more than 2$\boldsymbol\times$ reduced complexity.
\end{abstract}

% Massive multi-user multiple-input multiple-output (MU-MIMO) wireless systems operating at millimeter-wave (mmWave) frequencies enable simultaneous wideband data transmission to a large number of users. In order to reduce the complexity of MU precoding in all-digital basestation architectures, we propose a two-stage precoding architecture that first performs precoding using a sparse matrix in the beamspace domain, followed by an inverse fast Fourier transform that converts the result to the antenna domain. The sparse precoding matrix requires a small number of multipliers and enables regular hardware architectures, which allows the design of hardware-efficient all-digital precoders. Simulation results demonstrate that our methods approach the error-rate of conventional Wiener filter precoding with more than 2x reduced complexity.
%
	
% !TEX root = 21COMLET_beampre.tex
% DO NOT REMOVE THE ABOVE COMMENT!

\section{Introduction} 
\label{sec:intro}

Massive multi-user (MU) multiple-input multiple-output (MIMO) systems operating at millimeter-wave (mmWave) frequencies enable simultaneous, wideband wireless transmission to a large number of user equipments (UEs)~\cite{larsson14a,rappaport13a}.
While the large contiguous bandwidths available at mmWave {frequencies} enable high per-UE data rates, the strong atmospheric absorption necessitates MU precoding to provide sufficiently high signal-to-noise ratios (SNRs) at the UE side. 
Since massive MU-MIMO equips the infrastructure basestations (BSs) with a large number of antennas, fine-grained beamforming and  simultaneous data transmission to multiple UEs via spatial multiplexing is possible.
Hybrid analog-digital beamforming architectures for mmWave systems have been proposed in~\cite{alkhateeb14a,gao16,el-ayach14a}. 
{However, recent results in \cite{yan2019performance,roth18a,skrimponis20206gsummit} suggest that all-digital architectures enable} superior beamforming and spatial multiplexing capabilities, while achieving comparable system costs and {radio-frequency (RF)} power consumption by deploying low-precision data converters. 
In order to successfully deploy all-digital BS architectures in practice, novel {hardware- and power-efficient} baseband processing algorithms for channel estimation, data detection, and MU precoding are necessary.

An emerging approach towards low-complexity baseband processing algorithms and simpler hardware architectures for all-digital BSs is to exploit beamspace sparsity~\cite{gao16bs,chen17bscs,SayeedGLOBECOM,mirfarshbafan19a,abdelghany2019precoding,seyedicassp20,mahdavi20beamspace}. 
Since mmWave propagation is highly directional, the UE signals arrive at the BS from only a few incident angles \cite{rappaport13a}. By taking a spatial discrete Fourier transform (DFT) across the antenna array (e.g., a uniform linear array), the received signal is transformed from the antenna domain to the beamspace domain, which concisely reveals the underlying angular sparsity~\cite{ alkhateeb14a,mo16b, LeeOMP}. 
The sparse nature of the received beamspace signals can then be exploited in order to design low-complexity baseband algorithms and more efficient hardware architectures \cite{gao16bs,chen17bscs,SayeedGLOBECOM,mirfarshbafan19a,seyedicassp20,abdelghany2019precoding,mahdavi20beamspace}.
In the uplink, beamspace data detectors have been proposed in \cite{seyedicassp20,mahdavi20beamspace}
and beamspace channel estimators in \cite{gao17bsdl,he18bsdl,mirfarshbafan19a,dai16bs}. In the downlink, MU beamspace precoders have been proposed only recently in~\cite{pal18bs,pal19,abdelghany2019precoding,SayeedGLOBECOM, tang_downlink_2016}.

\subsection{Contributions}
We propose two-stage beamspace precoding algorithms for all-digital mmWave massive MU-MIMO systems.
Our algorithms rely on orthogonal matching pursuit (OMP) to compute sparse precoding matrices in the beamspace domain, which can result in lower precoding complexity than conventional, linear antenna-domain precoders that perform a dense matrix-vector product. The precoded output is then converted to the antenna domain using an inverse fast Fourier transform (IFFT).  
We use simulations for mmWave channels to demonstrate that our algorithms approach the  bit error-rate (BER) performance of conventional, antenna-domain Wiener filter (WF) precoding, while reducing the complexity by more than 2$\times$.

\subsection{Notation}
Boldface lowercase and uppercase letters represent vectors and matrices, respectively.
For a vector $\bma$, the  $k$th entry is $a_k=[\veca]_k$. 
For a matrix $\bA$, the transpose is $\bA^\Tran$ and the conjugate transpose is $\bA^\Herm$; the $k$th column is $\bma_k = [\bA]_k$ and the $k$th row is $\udl{\bma}_k = [\bA^\Tran]_k^\Tran$. 
For an index set~$\Omega$, $\bA_\Omega$ refers to the submatrix of $\bA$ with columns taken from $\Omega$. 
The $\ell_2$-norm of $\bma$ is $\|\veca\|$, the number of nonzero entries of $\bma$ is denoted by $\|\veca\|_0$, 
and the Frobenius norm of $\bA$ is $\|\bA\|_F$.
The $N\times N$ identity matrix is $\bI_N$ and the $N\times M$ all-zeros matrix is $\mathbf{0}_{N\times M}$. The $N\times N$ unitary DFT matrix is~$\bF_N$. 
The unit vector $\bme_n$ contains a $1$ in the $n$th entry and zeros otherwise.
Vectors and matrices in the beamspace domain are denoted with a bar, e.g., $\bar\bma$ and $\bar\bA$.
The set of integers $\{1,\dots,N\}$ is~$\setint{N}$.
%

% !TEX root = 21COMLET_beampre.tex
% DO NOT REMOVE THE ABOVE COMMENT!

\section{mmWave Massive MU-MIMO Downlink }
\label{sec:systemmodel}

\subsection{Downlink Channel and System Model}

We consider the mmWave massive MU-MIMO downlink, in which a BS with a $B$-antenna uniform linear array (ULA) transmits data to $U$ single-antenna\footnote{{With linear receive-side combining, the case of multiple-antenna receivers can be reduced to the single-antenna model as linear combinations of sparse channel vectors in beamspace typically remain to be sparse.}}
For illustrative purposes only, we model wave propagation from the BS to UE $u$ with the standard plane-wave approximation~\cite{tse05a}
$ {\udl{\vech}}_u = \sum_{\ell=0}^{L-1} \alpha_\ell \udl{\veca}(\phi_\ell)$,
where $L$ refers to the number of transmission paths between UE~$u$ and the BS antenna array (including a possible LoS path), 
$\alpha_l\in\opC$ is the complex-valued channel gain of the $\ell$th transmission path, and  
\begin{align} \label{eq:complexsinusoids}
\udl{\veca}(\phi_\ell) = \big[1, e^{j\phi_\ell},e^{j2\phi_\ell},\dots, e^{j(B-1)\phi_\ell} \big],
\end{align}
where $\phi_\ell$ is the spatial frequency determined by the $\ell$th path's incident angle to the ULA.
The downlink channel matrix  $\bH\in\complexset^{U\times B}$ comprises the rows ${\udl{\bmh}}_u$ for $u\in\setint{U}$.
In \fref{sec:results}, we show simulation results with more realistic mmWave channel vectors that do not rely on the plane-wave approximation, generated from the mmMAGIC QuaDRiGa model~\cite{QuaDRiGa}.

We consider a block-fading frequency-flat channel, in which the channel stays constant over a block of $T$ time slots. We model the downlink input-output relation as follows:
\begin{align} \label{eq:downlinkmodel}
\bmy = \bH \bmx + \bmn.
\end{align}
Here, the $U$-dimensional vector $\bmy \in \mathbb{C}^U$ comprises the signals received at all $U$ UEs
and the entries of the noise vector $\bmn\in\complexset^B$ are i.i.d.\ circularly-symmetric complex Gaussian with (known) variance~$N_0$.
To mitigate MU interference, the BS must precode the transmit symbols. 
To this end, a $B$-dimensional antenna-domain precoded vector $\bmx$ is formed according to  
\begin{align} \label{eq:precodingfunction}
\bmx=\setP(\bms,\bH,\No,\rho^2),
\end{align} 
where the transmit vector $\bms \in \setO^U$ contains the $U$ data symbols to be transmitted to the UEs, 
$\setO$ is the constellation set (e.g., 16-QAM), the transmit signals are assumed to be i.i.d.\ zero-mean and normalized so that $\Ex{}{|s_u|^2}=\Es$ for all $u\in\setint{U}$ and~$\rho^2$ is the average power constraint so that
$\Ex{\bms}{\|\bmx\|^2}\leq \rho^2$.

\subsection{MSE-Optimal Linear Precoding}
\label{sec:linprecoding}
To minimize the precoding complexity, we focus on linear precoders for which the precoding rule in \eqref{eq:precodingfunction} is linear, i.e., 
\begin{align} \label{eq:linearprecoding}
\bmx=\setP(\bms,\bH,\No,\rho^2)=\bP \bms
\end{align}
with the precoding matrix $\bP \in \complexset^{B\times U}$.
Since multi-antenna transmission causes an array gain, each UE $u$ performs scalar equalization of the received signal $y_u$ with a precoding factor $\beta_u\in\complexset$ according to
$\hat s_u = \beta_u y_u$, $u=1,\ldots,U$. 
As in \cite{li2018feedforward}, we consider pilot-based estimation of the precoding factors: In the first time slot, the BS transmits $U$ pilots with energy $\Es$, which are then used at each UE to estimate~$\beta_u$. 

We focus on linear precoders that minimize the UE-side mean-square error (MSE) for a common $\beta\in\complexset$ so that
\begin{align} 
\textit{MSE} & \define \Ex{\bms,\bmn}{\|\bms-\hat\bms\|^2} = \Ex{\bms,\bmn}{\|\bms-\beta\bmy\|^2} \\
& =  \Ex{\bms}{\|\bms-\beta \bH\bmx\|^2} + |\beta|^2 U \No \label{eq:mse}
\end{align}
is minimized.
The MSE-optimal linear precoder is known as the Wiener filter (WF) precoder~\cite{joham05a},
where the precoding matrix $ \bP^{\text{WF}} = \bQ^{\text{WF}}/\beta(\bQ^{\text{WF}})$ is given by 
\begin{align}
\label{eq:originalQmatrix}
\bQ^{\text{WF}} = \left(  \bH^H \bH + \kappa^{\text{WF}} \bI_B \right )^{-1} \bH^H.
\end{align}
Here, $\kappa^{\text{WF}} =  {U\No}/{\rho^2}$, and {$\beta: \opC^{B\times U}\to \opR$} is a function that computes a pre-factor to satisfy the power constraint:
\begin{align}
\label{eq:precodingfactor}
\beta(\bQ) = \textstyle \sqrt{{\tr\big(\bQ^H\bQ\big) \Es}/{\rho^2}} .  
\end{align}

As it will become useful later, one can alternatively obtain the (unnormalized) WF precoding matrix $\bQ^{\text{WF}}$ in \eqref{eq:originalQmatrix} by solving the following unconstrained optimization problem \cite{castaneda19fame}:
\begin{align}\label{eq:WF}
\bQ^{\text{WF}} = \argmin_{\bQ\in\complexset^{B\times U}} \,\, \|\bH\bQ-\bI_U\|_F^2 + \kappa^{\text{WF}} \|\bQ\|_F^2.
\end{align}
\subsection{Linear Precoding in the Beamspace Domain}
\label{sec:beamspacepc}
In order to reduce the complexity of conventional, antenna-domain WF precoding $\bmx=\bP^{\text{WF}}\bms$, one can perform linear precoding in the beamspace domain \cite{abdelghany2019precoding}.
The key idea is to deploy linear precoders of the following form: 
\begin{align} \label{eq:linearbeamspaceprecoding}
\bmx=\setP(\bms,{\bar\bH},\No,\rho^2)=\bF_B^\Herm {\bar\bP} \bms.
\end{align} 
Here, $\bar\bH = \bH \bF_B$ is the beamspace representation of the mmWave MIMO channel matrix. 
Since the rows of $\bH$ consist of a superposition of a few complex-valued sinusoids, e.g., as in \eqref{eq:complexsinusoids}, the rows of $\bar\bH$ are  sparse~\cite{alkhateeb14a,sayeed2014precoder,alkhateeb15a,LeeOMP} and large entries correspond to the strong transmission paths, i.e., the beams for each user. 
This property enables one to design beamspace precoding matrices $\bar\bP$ with sparse columns, in which the nonzero entries correspond to the selected beams from the rows of $\bar\bH$. If the number of beams is proportional to~$U$, then we can capture the beams that carry the information of all users. 
We then compute the beamspace-domain precoding vector $\bar\bmx=\bar\bP\bms$, which requires lower complexity than~\fref{eq:linearprecoding} due to fewer nonzero multiplications. Finally, we convert $\bar\vecx$ into the antenna domain using an IFFT as in~\fref{eq:linearbeamspaceprecoding}.
% !TEX root = 21COMLET_beampre.tex
% DO NOT REMOVE THE ABOVE COMMENT!

\section{Sparse Beamspace Precoding Algorithms} \label{sec:algorithms}
We now propose algorithms to compute sparse precoding matrices that are suitable for beamspace precoding as in~\eqref{eq:linearbeamspaceprecoding}.
We start by an OMP-based algorithm, and then propose alternative algorithms with additional structure on the sparse matrix {$\bar\bP$}, which simplify corresponding hardware architectures.

\subsection{Sparse Beamspace Precoding (SBP)} %OMP
\label{sec:sbp}
In order to design SBP matrices, we modify the optimization problem in \eqref{eq:WF} to deliver sparse matrices. 
{Our algorithms do not guarantee that the solution to a sparsity-constrained version of (9) is equal to that of the sparsity-constrained MSE minimization problem.
However, our results show that our methods lead to solutions with small MSE.}
As a first method, we propose to solve the following optimization problem 
\begin{align} \label{eq:matrixproblemsparse}
\bar\bQ^{\text{SBP}} = 
\argmin_{\bar\bQ\in\mathfrak{S}_{\text{SBP}}} \,\, \|\bar\bH\bar\bQ-\bI_U\|_F^2 + \kappa^{\text{WF}} \|\bar\bQ\|_F^2 .
\end{align}
where we impose a constraint that ensures each column of $\bar\bQ$ to have exactly $K$ {nonzero} entries, i.e., 
\begin{align}
\mathfrak{S}_\SBP \triangleq \{ \bar\bQ \in\complexset^{B\times U} :  \|\bar\bmq_u\|_0 = K,  u=1,\ldots,U\}.
\end{align}
We then normalize the matrix $\bar\bQ^{\text{SBP}}$ to obtain the SBP matrix $\bar\bP^{\text{SBP}}= \bar\bQ^{\text{SBP}}/\beta(\bar\bQ^{\text{SBP}})$, where $\beta(\bar\bQ^{\text{SBP}})$ was defined in \eqref{eq:precodingfactor}.
It is important to realize that one can solve the problem in~\eqref{eq:matrixproblemsparse} on a per-column basis, i.e., we can solve 
\begin{align} \label{eq:sparseapproximation}
\bar\bmq^{\text{SBP}}_u=
\argmin_{\bar\bmq\in\complexset^{B},\|\bar\bmq\|_0 = K} &  \|\bar\bH\bar\bmq-\bme_u\|^2 + \kappa^{\text{WF}} \|\bar\bmq\|^2 
\end{align}
for $u=1,\ldots,U$. Unfortunately, this sparse approximation problem is NP-hard \cite{natarajan0norm} and thus must be solved using approximate methods. We propose to compute an approximate solution to \eqref{eq:sparseapproximation} using OMP \cite{omp}, as detailed next.
We note that the iterative algorithms detailed below make locally optimal decisions in every iteration, without any guarantees that the final solution will be globally optimal.

Let ${\bar\bmq}_u^{(k)}\in\opC^{k}$ be the vector computed after the $k$th OMP iteration, and ${\bar\bmr}_u^{(k)}$ the associated residual.
Let $\Omegac_u^{(k)}$ be the set of indices of the $k$ nonzero entries of $\bar\bmq_u$,
and let $\Omega_u^{(k)}$ be the set of available indices for the new nonzero entry in the $(k+1)$th iteration. Here, $\Omega_u^{(k)}= \setint{B} \setminus \Omegac_u^{(k)},\forall k$.
We initialize the available and already-selected indices $\Omega^{(0)}_u = \setint{B}$, $\Omegac^{(0)}_u =\varnothing$, and the residual $\bar\bmr^{(0)}_u=\bme_u$. Then, repeat the following three steps for iterations $k=1,\ldots,K$:
(i) Identify the next best 
beam index by correlating the residual with the columns of $\bar\bH$,
\begin{align} \label{eq:selection_criterion}
b_u^{(k)}=\argmax_{b\in\Omega_u^{(k-1)}} \, |\bar\bmh_b^\Herm \bar\bmr_u^{(k-1)}|,
\end{align}
and augment the support set, $\Omegac^{(k)}_u = \Omegac^{(k-1)}_u \cup \{b^{(k)}_u\}$. By definition, $b^{(k)}_u$ is unavailable for selection in subsequent iterations and we use $\Omega^{(k)}_u = \Omega^{(k-1)}_u \setminus \{b^{(k)}_u\}$.
(ii) Update the SBP vector as for the WF precoder, 
\begin{align}
\bar\bmq_u^{(k)}=(\bar\bH^\Herm_{\Omegac_u^{(k)}} \bar\bH_{\Omegac_u^{(k)}}+ \kappa^{\text{WF}} \bI_k)^{-1} \bar\bH^\Herm_{\Omegac_u^{(k)}} \bme_u.
\end{align}
(iii) Update the residual, $\bar\bmr_u^{(k)}=\bme_u - \bar\bH_{\Omegac_u^{(k)}} \bar\bmq_u^{(k)}$.
After $K$ iterations, {the entries of  $\bar\bmq_u^{(K)}$ are assigned to $[\bar\bmq_u]_b,b\in\Omegac^{(K)}$, i.e., the nonzero entries of the SBP column $\bar\bmq_u$;} this procedure is repeated for all columns $\bar\bmq_u$, $u\in\setint{U}$, of the unnormalized SBP matrix $\bar\bQ^{\text{SBP}}$. 
We then normalize the sparse matrix $\bar\bQ^{\text{SBP}}$ to obtain the SBP matrix $\bar\bP^{\text{SBP}}= \bar\bQ^{\text{SBP}}/\beta(\bar\bQ^{\text{SBP}})$, {where the precoding factor is given by~\eqref{eq:precodingfactor}.}
The resulting SBP matrix~$\bar\bP^{\text{SBP}}$ contains, as desired, exactly $KU$ nonzero entries.
\subsection{Row-Select Sparse Beamspace Precoding (RS)}
\label{sec:rowmax}
Although the above approach results in a sparse precoding matrix with $KU$ nonzero entries, the unstructured nature of the nonzero entries in $\bar\bP$ prevents efficient, parallel hardware architectures that perform the sparse matrix-vector multiplication at high rates. 
To overcome this issue, we propose to enforce \emph{structured} sparsity in the matrix $\bar\bP$ such that the rows have either all ($U$) nonzero entries or all zeros, so we can only store the nonzero rows and use efficient hardware for the sparse matrix-vector multiplication. 
Concretely, we aim to solve the precoding problem in~\eqref{eq:matrixproblemsparse} with the constraint set 
\begin{align} \label{eq:shitset2}
\mathfrak{S}_{\text{RS}} \triangleq \textstyle \Big\{ \bar\bQ \in\complexset^{B\times U} : \, &  \|\udl{\bar\bmq}_b\|_0 = \begin{cases} U,\ \text{if $b$ is selected}\\ 0,\ \text{otherwise} \end{cases}\!\!\!\!\!\!, \notag \\
& \|\bar\bmq_u\|_0 = K,  u=1,\ldots,U \Big\},
\end{align}
which requires us to find $K$ nonzero rows of the unnormalized precoding matrix $\bar\bQ$, each with $U$ nonzero entries.
This problem resembles a multiple measurement vector (MMV) problem~\cite{mmv} and we use an OMP-MMV-like algorithm; we call the method  \textit{Row-Select SBP}, simply denoted by RS. 

Let $\Omegac^{(k)}$ denote the rows of $\bar\bQ$ that are
selected as nonzero in the first $k$ iterations, 
and $\Omega^{(k)} = \setint{B} \setminus \Omegac^{(k)}$ the remaining ones,
i.e., rows available for selection in the $(k+1)$th iteration.
Let $\bar\bQ^{(k)}\in\opC^{k\times U}$ denote a submatrix of the precoding matrix computed at the $k$th iteration,
and $\bar\bR^{(k)}$ the residual.
We initialize the set of selected nonzero rows $\Omegac^{(0)} = \varnothing$
and the residual $\bar\bR^{(0)}=\bI_U$. 
We repeat the following steps for iterations $k=1,\ldots,K$: 
(i) Identify the next best beam index,
\begin{align}
\label{eq:rowmaxselect}
    \hat{b}^{(k)}=\argmax_{b\in \Omega^{(k-1)}} \|\bar\bmh_b^\Herm \bar\bR^{(k-1)}\|_2,
\end{align}
and add this index to the support set $\Omegac^{(k)} = \Omegac^{(k-1)} \cup \{\hat{b}^{(k)}\}$.
By definition, $\Omega^{(k)} = \Omega^{(k-1)} \setminus \{\hat{b}^{(k)}\}$.
(ii) Update the submatrix of the precoding matrix, 
\begin{align}
\bar\bQ^{(k)}=(\bar\bH^\Herm_{\Omegac^{(k)}}\bar\bH_{\Omegac^{(k)}}+ \kappa^{\text{WF}} \bI_k)^{-1} \bar\bH^\Herm_{\Omegac^{(k)}} . 
\end{align}
(iii) Update the residual, $\bar\bR^{(k)}=\bI_U - \bar\bH_{\Omegac^{(k)}} \bar\bQ^{(k)}.$
After $K$ iterations, the rows of $\bar\bQ^{(K)}$ deliver the nonzero rows $\udl{\bar\bmq}_b,b\in \Omegac^{(K)}$, of the unnormalized RS matrix $\bar\bQ^{\text{RS}}$, which has exactly $KU$ nonzero entries 
with $\udl{\bar\bmq}_b$ containing exactly $U$ nonzeros.
The RS matrix is obtained by $\bar\bP^{\text{RS}}= \bar\bQ^{\text{RS}}/\beta\big(\bar\bQ^{\text{RS}}\big)$ {with the normalization factor given by~\eqref{eq:precodingfactor}.}

\subsection{Simplified One-Shot SBP {and RS} Algorithms}
\label{sec:1S}
All of the above methods require $K$ iterations to construct $K$-sparse beam vectors for each UE. 
To further reduce the preprocessing complexity, we propose simplified methods that require only one iteration. 
For the counterpart of SBP, we construct the support set $\Omega_u$ per user $u$ by selecting $K$ beam indices that maximize the criterion in \eqref{eq:selection_criterion}. 
For the counterpart of RS, we construct the support set of nonzero rows by selecting the $K$ beam indices maximizing \eqref{eq:rowmaxselect}.
We call each of these methods One-Shot SBP (1S-SBP) and One-Shot RS (1S-RS).

% !TEX root = 21COMLET_beampre.tex
% DO NOT REMOVE THE ABOVE COMMENT!

\section{Results}
\label{sec:results}
\subsection{Simulation Setup}
\label{sec:simsetup}
We simulate line-of-sight (LoS) {and non-LoS (nLoS)}
channel conditions, both including multiple reflective paths, using the QuaDRiGa mmMAGIC UMi model\cite{QuaDRiGa} at a carrier frequency of 60 GHz with $\lambda/2$-spaced antennas arranged as a ULA.
We generate channel matrices for a mmWave massive MIMO system with $B=128$ antennas and $U = 16$ UEs.
The UEs are placed randomly in a $120^\circ$ circular sector around the BS between a distance of 25\,m and 112\,m, and we assume a minimum UE separation of  $1^\circ$.
We assume UE-side power control so that the norms of the UE's channel vectors differ by at most  $6$\,dB. 
To account for channel estimation errors in the uplink, we assume that the BS has access to a noisy version of $\bH$ modeled as
   $\hat\bH = \sqrt{1-\epsilon}\bH + \sqrt{\epsilon}\bZ$
as in~\cite{jacobsson17d}. Here, $\bZ\sim\mathcal{CN}(\mathbf{0}_{U\times B},\bI_N)$ models the  error for pilot-based channel estimation in the uplink and we set $\epsilon = 0.0099$ so that the channel estimation error corresponds to operating the system at $20$\,dB SNR. 
In our simulations, we use the beamspace channel estimation (BEACHES) algorithm from~\cite{mirfarshbafan19a}, 64-QAM transmission, and UE-side hard-output data detection.

We simulate the uncoded BER versus the normalized transmit power $\rho^2/\No$ for the sparsity parameters $K=U$ and $K=2U$ using the proposed precoders from \fref{sec:algorithms}. 
Here, we choose $K$ proportional to $U$ to capture the beams for all users, while also aiming to keep $K$ as small as possible to minimize complexity.
As baseline methods, we simulate the performance of the WF precoder from \fref{sec:linprecoding} and maximum ratio transmission (MRT). 
We also compare with the algorithms in~\cite{abdelghany2019precoding}, \cite{pal18bs}, and \cite{pal19}, referred to as local WF, QR, and greedy beam selection (GBS), respectively.
Local WF \cite{abdelghany2019precoding} approximates the beamspace channel vectors by preserving the $K$-sized window of $\udl{\bar\vech}_u$ with the highest energy and setting the remaining entries to zero.
To enable a fair comparison, the precoding coefficients are selected to minimize the MSE as in~\fref{eq:mse}, whereas the original objective in \cite{abdelghany2019precoding} maximizes the minimum UE-side SINR. 
This algorithm requires the inversion and multiplication of sparse matrices, but as sparsity is not explicitly imposed, there is no guarantee on the number of zeros in the resulting precoding matrix. 

Regarding QR~\cite{pal18bs} and GBS~\cite{pal19}, these algorithms originally pick $K=U$ beams, whereas we vary $K$ for fair~comparison with our algorithms.
For GBS, we implement this modification by repeating the per-user beam allocation process $K/U$ times.

\subsection{Complexity Analysis} 
\label{sec:complexity}
We provide a complexity analysis in~\fref{tbl:complexity}, in which we summarize the number of real-valued multiplications required during preprocessing (calculating the precoding matrix) and precoding (applying the precoding matrix to $T$ transmit vectors), following the analysis in \cite{castaneda19fame}. As in \cite{seyedicassp20}, we assume a complexity of $2B\log_2B$ for a $B$-point (I)FFT.
Since RS has the same total complexity as SBP, SBP represents both methods;  the same holds for 1S-SBP and 1S-RS. 
For local WF, $M$ stands for the average number of nonzeros in the precoding matrix based on experiments, where we assume a zero entry if the absolute value is smaller than $10^{-7}$.
For the QR and GBS methods, we assume a Householder QR factorization~\cite{golub}.
We note that $K$ should be larger for less sparse channels, which increases the complexity of all sparsity-exploiting algorithms.
\begin{table}[tbp]
\centering
\caption{Complexity of various precoding methods.}
\label{tbl:complexity}	
\hspace{-1.95cm}
\resizebox{0.75\columnwidth}{!}{
\begin{minipage}[c]{0.95 \columnwidth}
		\begin{tabular}{@{}lll@{}}
			\toprule
			%\cmidrule{2-5}  \cmidrule{7-10} 
			{Algorithm}\!\!\! & {Preprocessing complexity} \!\!\!& {Precoding complexity}\\
			\midrule
			{WF} & $2U^3 + 6BU^2 - 2U(U+1)+1$ & $4TBU$\\
			{MRT} & 0 & $4TBU$\\
			{Local WF}\!\!\! & $2UB\log_2B + 2U^3 + 6KU^2 - 2U(U+1)+1$ & $4TMU \!+\! 2TB\log_2\!B$\\
			{QR} & $2UB\log_2\!B + 4\sum_{i=1}^U (i-1)(1+2(U-i))$\\	&$+12\sum_{i\!=\!0}^{B-K-1} (B-i)\sum_{j\!=\!0}^{U-1}(B\!-\!i\!-\!j\!)(U\!-\!j) $\\
			&$+ 4U^2+ 4K\sum_{i=0}^{U-1}(U-i)$ & $4TKU \!+\! 2TB\log_2\!B$\\
			{GBS} & $2UB\!\log_2B + 12\sum_{j\!=\!0}^{U-1}(K\!-\!j)(U\!-\!j) \!+\! 4U^2$\\
			&$+ 4\sum_{i=1}^U (i-1)(1+2(U-i)) + 4K\sum_{i=0}^{U-1}(U-i)\!\!\!\!$ & $4TKU \!+\! 2TB\log_2\!B$\\
			{SBP} & $2UB\log_2B + 4KB(U+2) + 2UK(K+1)$\\			
			& $+ 2\sum_{k=1}^K (k^3+3Uk^2-(U+1)k+1) $ & $4TKU \!+\! 2TB\log_2\!B$ \\
			{1S-SBP} & $2UB\log_2\!B$\\ &$+U(4B(U+2)+2K^3+6UK^2-2(U+1)K+1)$ & $4TKU \!+\! 2TB\log_2\!B$\\
			\bottomrule 
		\end{tabular}		
	\end{minipage}}
\vspace{-0.2cm}	
\end{table}

In \fref{fig:complexity}, we show the speed-up of the algorithms compared to MRT, which we define as the ratio of the total complexity required by MRT to that of the algorithm, with respect to the number of transmissions~$T$ within a channel coherence interval. 
For $T\to\infty$, the asymptotic speed-up of our algorithms is $\gamma\triangleq \frac{2BU}{B\log_2B+2UK}$.
\fref{fig:complexity} reveals that QR is the most complex method. 
For a small coherence time $T$, 
GBS and WF are less complex than our algorithms, but WF could be the most preferable given that it achieves the smallest MSE. 
Our SBP-based methods catch up with the speed of GBS as $T$ increases, and $T$ can be as large as $10^5$~\cite{seyedicassp20} in practical mmWave systems. 
We see that already for $T>10^3$, 1S-SBP is up to 2.91$\times$ faster than MRT. 
SBP requires larger $T$ and smaller $K$ than 1S-SBP to outperform the baseline methods. 
\subsection{Bit Error-Rate Performance}
\fref{fig:berplot} shows the uncoded BER for the scenarios in \fref{sec:simsetup} for $B=128$ BS antennas, $U=16$ users. We consider two sparsity levels $K=16$ (a,c) and $K=32$ (b, d) under LoS (a,b) and nLoS (c,d) conditions.
To compare these algorithms, we consider a target BER of $2\%$.
In the LoS scenario, SBP, RS, and 1S-SBP outperform local WF, GBS and MRT. 
QR has a similar BER performance to our algorithms, but it is not preferable as its complexity is much higher than WF as shown in \fref{sec:complexity}. 
For $K=U$, \fref{fig:los_u} shows that the SNR required by SBP to achieve the target BER is $1.5$\,dB higher than WF.
In \fref{fig:los_2u}, the BER of all our methods approach to that of WF. Here, the one-shot variants are the most preferable as they have lower complexity than the iterative methods, while performing similarly in BER.
In the nLoS scenario of \fref{fig:nlos_u}, as the channel is less sparse than in the LoS case, we observe that $K=U$ is not sufficient for any of the SBP methods to perform comparably to WF. 
Moreover, SBP performs worse than 1S-SBP, which exemplifies a case of our iterative algorithms leading to globally suboptimal solutions.
For $K=2U$, \fref{fig:nlos_2u} shows that the SNRs required by SBP and RS to achieve the target BER are $1.5$\,dB higher than~WF.
The one-shot versions do not perform well in BER even for $K=2U$. Hence, to obtain comparable BER performance to WF, our iterative SBP algorithms are preferred over the one-shot variants if the channel vectors are less sparse.

\newcommand{\figsize}{0.237}
\begin{figure}[tp]
 \vspace{-0.15cm}
    \centering
   \subfloat[$U=16,\, K=16,\gamma=2.91$]
	{
    \includegraphics[width=\figsize\textwidth]{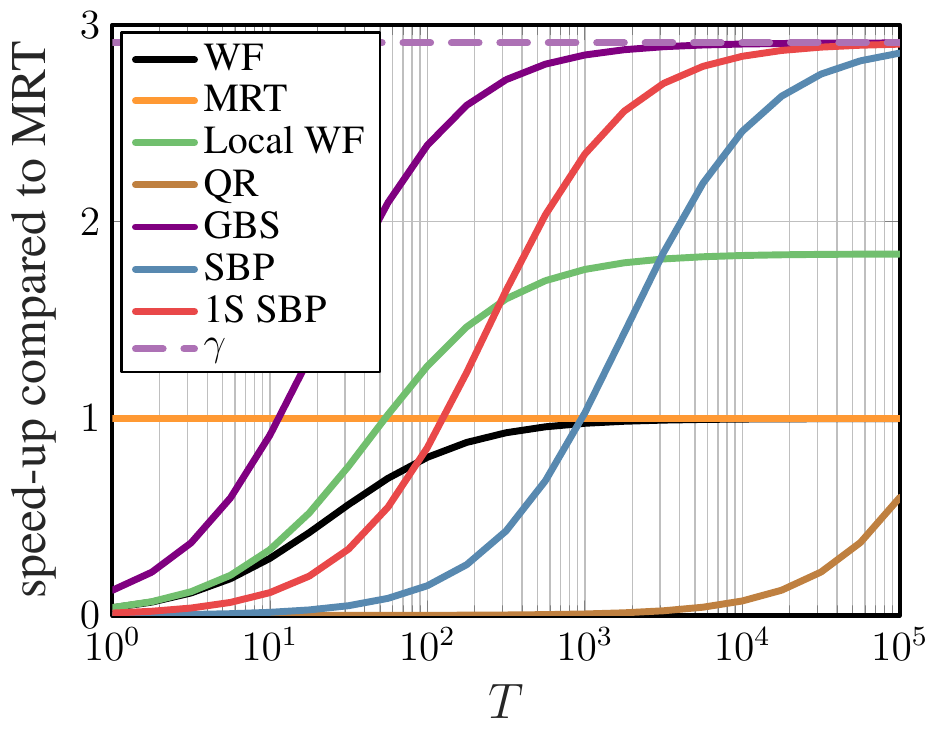}
    }
    \subfloat[$U=16,\, K=32,\gamma=2.13$]
	{
    \includegraphics[width=\figsize\textwidth]{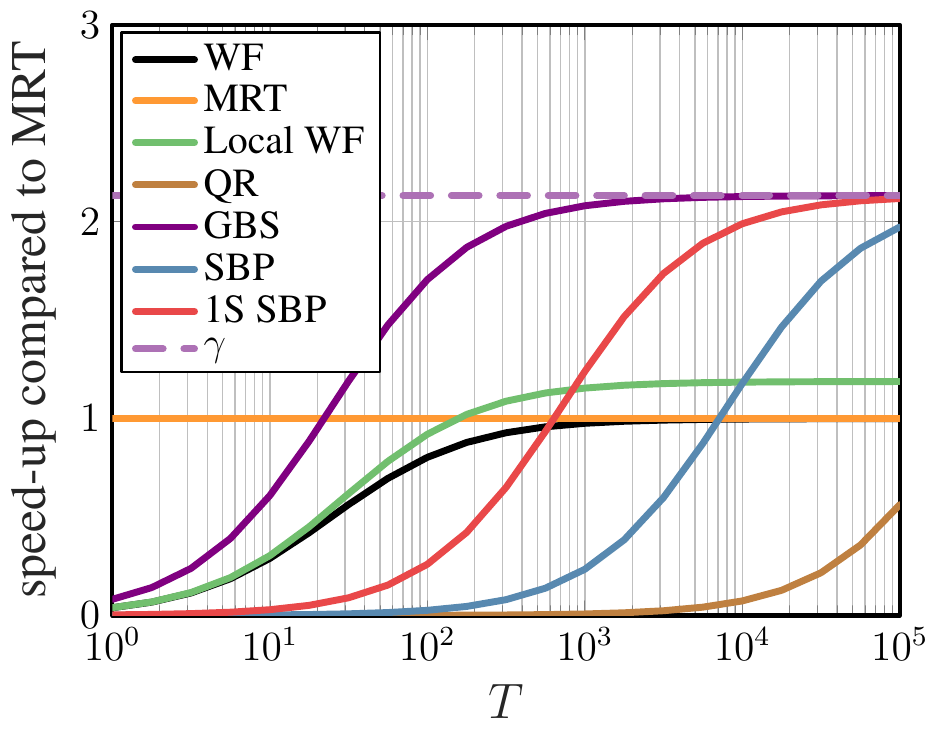}
    }
    \caption{Speed-up compared to MRT vs the number of transmissions ($T$)  evaluated by the number of real-valued multiplications for $B = 128$ BS antennas, $U=16$ users, and for sparsity levels $K=U$ and $K=2U$. Our sparse beamspace precoding algorithms are up to 2.91$\times$ faster than MRT.}
    \label{fig:complexity}
    \vspace{-0.3cm}
\end{figure}

\begin{figure*}[tp]
    \centering
    \subfloat[LoS, $U=16,\, K=16$]
    {
    \includegraphics[width=\figsize\textwidth]{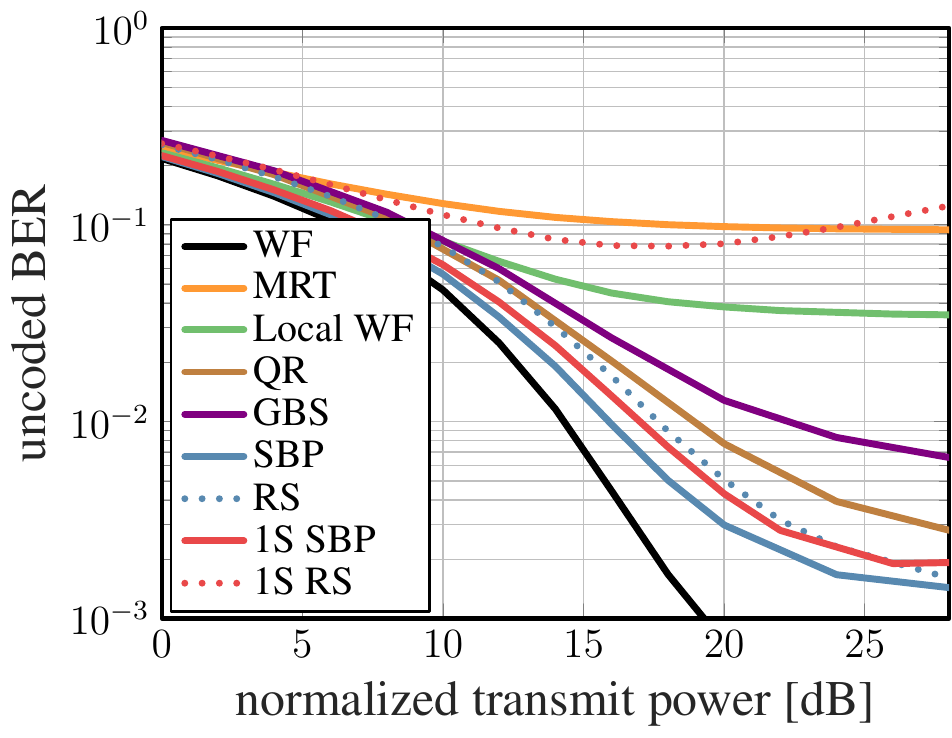} \label{fig:los_u}
    }
    \subfloat[LoS, $U=16,\, K=32$]
	{
    \includegraphics[width=\figsize\textwidth]{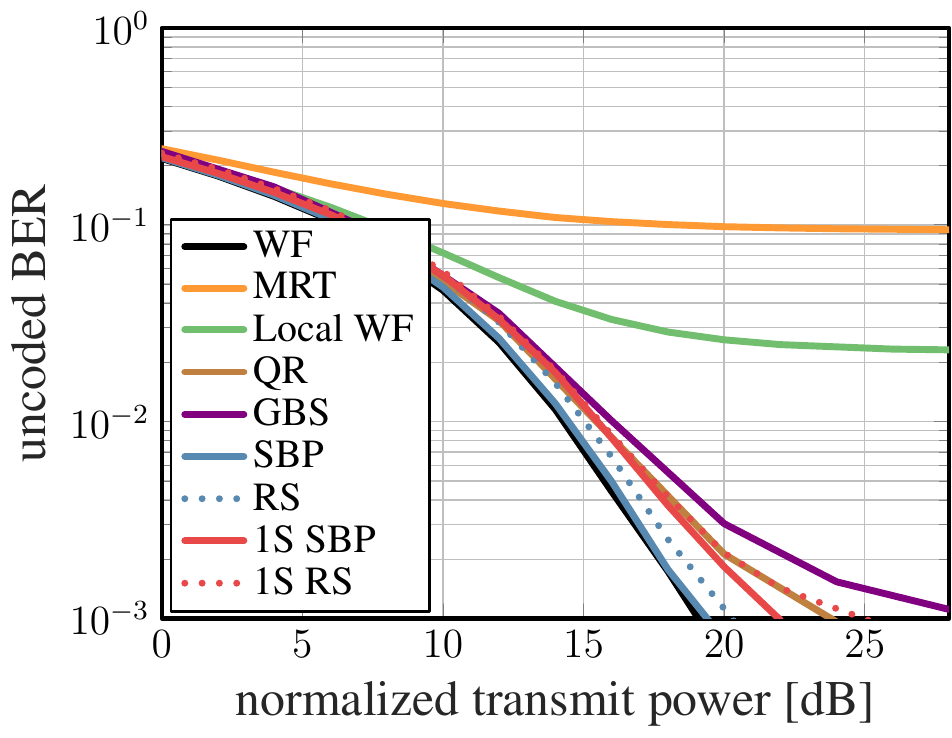}\label{fig:los_2u}
    }
    \subfloat[nLoS, $U=16,\, K=16$]
	{
    \includegraphics[width=\figsize\textwidth]{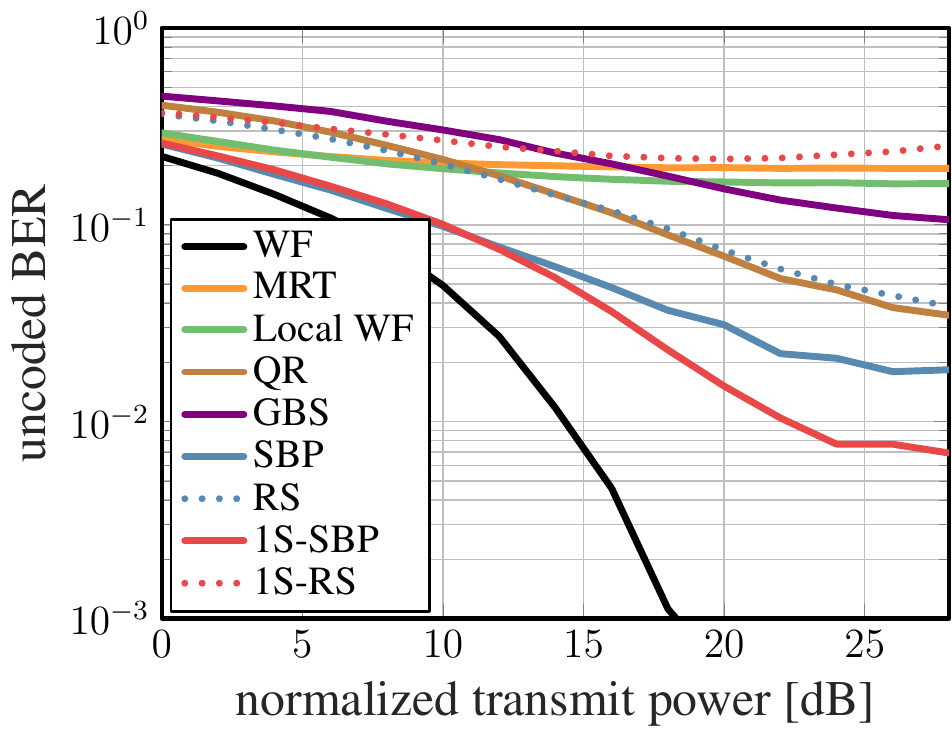} \label{fig:nlos_u}
    }
    \subfloat[nLoS, $U=16,\, K=32$]
	{
    \includegraphics[width=\figsize\textwidth]{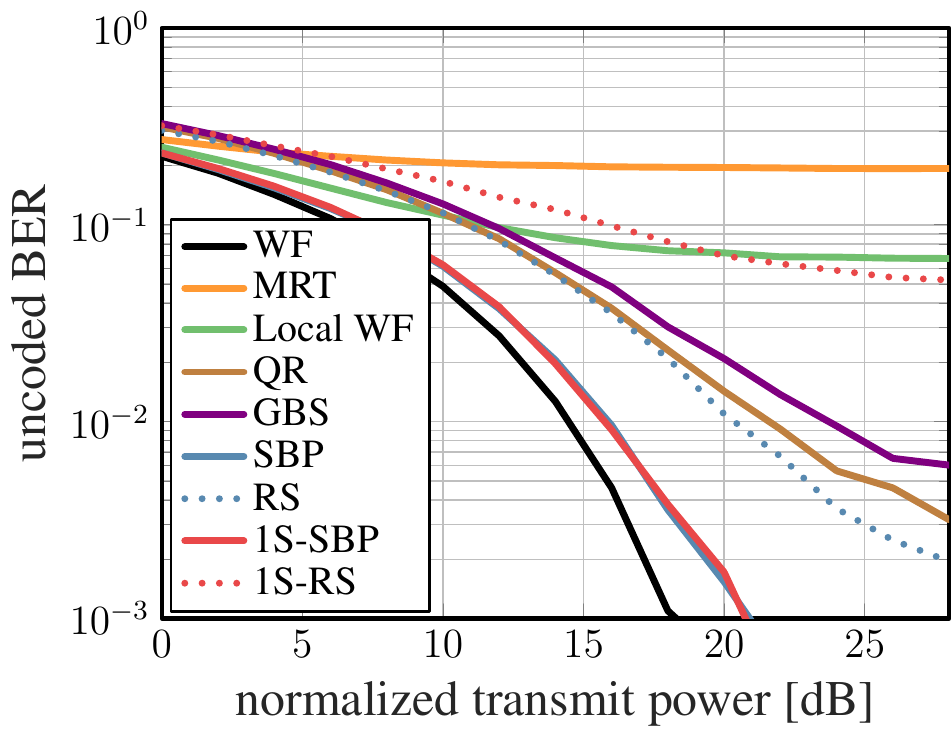} \label{fig:nlos_2u}
    }
    \vspace{-0.0cm}
    \caption{Bit error rate (BER)  results of LoS (a,b) and nLoS (c,d) scenarios with $B=128$ BS antennas, $U=16$ users, and for sparsity levels $K=U$ and $K=2U$. The proposed sparse beamspace precoding algorithms are able to achieve a performance close to Wiener filter (WF) for sparsity level $K=2U$.}
    \label{fig:berplot}
      \vspace{-0.3cm}
\end{figure*}

% !TEX root = 21COMLET_beampre.tex
% DO NOT REMOVE THE ABOVE COMMENT!

\section{Conclusions}
We have proposed four different algorithms to perform sparse precoding in the beamspace domain. Our algorithms consist of two stages: 
The first stage performs sparse beamspace precoding; the second stage converts the precoded vector to the antenna domain using fast Fourier transform. 
{Our simulation results for LoS and nLoS mmWave massive MU-MIMO channels have shown that our sparse beamspace precoding algorithms reduce the complexity by more than $2\times$ compared to traditional, antenna-domain Wiener filter precoding while delivering comparable error-rate performance.
A hardware implementation of our algorithms is part of ongoing work.}

\balance
\bibliographystyle{IEEEtran}
\bibliography{bib/VIPabbrv,bib/publishers,bib/confs-jrnls,bib/REFs,vipbib}
\balance
	
\end{document}